 \documentclass[twocolumn]{aastex61}

\newcommand\aastex{AAS\TeX}

\shorttitle{\aastex\ ICL in the Fornax Cluster}
\shortauthors{Iodice et al.}

\begin{document}

\title{Intra-cluster patches of baryons in the core of the Fornax cluster}

\correspondingauthor{Enrichetta Iodice}
\email{iodice@na.astro.it}

\author[0000-0003-4291-0005]{E. Iodice}
\affil{INAF-Astronomical Observatory of Capodimonte, via Moiariello 16, Naples, I-80131, Italy}

\author{M. Spavone}
\affiliation{INAF-Astronomical Observatory of Capodimonte, via Moiariello 16, Naples, I-80131, Italy}

\author{M. Cantiello}
\affiliation{INAF-Astronomical Observatory of Teramo, Via Maggini, 64100, Teramo, Italy}

\author{R. D'Abrusco}
\affiliation{Smithsonian Astrophysical Observatory/Chandra X-ray centre, 02138 Cambridge (MA), US}

\author{M. Capaccioli}
\affiliation{University of Naples ``Federico II'', C.U. Monte SantÕAngelo, Via Cinthia, 80126, Naples, Italy}

\author{M. Hilker}
\affiliation{European Southern Observatory, Karl-Schwarzschild-Strasse 2, D-85748 Garching bei Munchen, Germany}

\author{S. Mieske}
\affiliation{European Southern Observatory, Alonso de Cordova 3107, Vitacura, Santiago, Chile}

\author{N.R. Napolitano}
\affiliation{INAF-Astronomical Observatory of Capodimonte, via Moiariello 16, Naples, I-80131, Italy}

\author{R. F. Peletier}
\affiliation{Kapteyn Astronomical Institute, University of Groningen, PO Box 72, 9700 AV Groningen, The Netherlands}

\author{L. Limatola}
\affiliation{INAF-Astronomical Observatory of Capodimonte, via Moiariello 16, Naples, I-80131, Italy}

\author{A. Grado}
\affiliation{INAF-Astronomical Observatory of Capodimonte, via Moiariello 16, Naples, I-80131, Italy}

\author{A. Venhola}
\affiliation{Division of Astronomy, Department of Physics, University of Oulu, Oulu, Finland}

\author{M. Paolillo}
\affiliation{Univ. of Naples ``Federico II'', C.U. Monte SantÕAngelo, Via Cinthia, 80126, Naples, Italy}

\author{G. Van de Ven}
\affiliation{Max Planck Institute for Astronomy, Heidelberg, Germany}

\author{P. Schipani}
\affiliation{INAF-Astronomical Observatory of Capodimonte, via Moiariello 16, Naples, I-80131, Italy}



\begin{abstract}
In the core of the Fornax cluster, on the West side of NGC 1399, we have detected a previously unknown region of intra-cluster light (ICL). It is made up by several faint ($\mu_r \simeq 28 - 29$~mag/arcsec$^2$) {\it patches} of diffuse light. The bulk of the ICL is located in between the three bright galaxies in the core, NGC~1387, NGC~1379 and NGC~1381, at $10\leq R \leq40$~arcmin ($\sim 58 - 230$~kpc) from the central galaxy NGC~1399. We show that the ICL is  the counterpart in the diffuse light of the known over-density in the population of blue globular clusters (GCs). 
The total $g$-band luminosity of the ICL is $L_g\simeq 8.3 \times 10^{9}$~$L_{\odot}$, which is $\sim5\%$ of the total luminosity of NGC~1399. This is consistent with the fraction of the blue GCs in the same region of the cluster.  
The ICL has  $g-r \sim 0.7$~mag, which is similar to the colors  in the halo of the bright galaxies in the cluster core. 
{  The new findings were compared with theoretical predictions for the ICL formation and they support a scenario in which the intra-cluster population detected in the core of the Fornax cluster is build up by the tidal stripping of material (stars and GCs) from galaxy outskirts in a close passage with the cD. Moreover, the diffuse form of the ICL and its location close to the core of the cluster is expected in a dynamically evolved cluster as Fornax.}

\end{abstract}




\section{Introduction} \label{intro}

Many studies have proved the existence of intra-cluster baryons, in stars and GCs, which are not gravitationally bound to any of the cluster members \citep[][]{Zibetti2005,Gonzalez2007,Arnaboldi2010,Tutukov2011,Mihos2015}.  The ICL is the diffuse component from the starlight that contributes to the surface brightness profile of the bright cluster galaxy (BCG) at magnitudes fainter than $\mu_R>26$~mag/arcsec$^2$ \citep{Zibetti2005,Mihos2005}. These are the regions of the stellar halos, where the ICL contribution increases with the distance from the center, being a part of the diffuse outer envelope \citep{Zibetti2005,Gonzalez2007,Iodice2016,Spavone2017}. 

Simulations suggest that, during the hierarchical mass assembly of the galaxy clusters,  the intra-cluster baryons  build up by the tidal stripping of material from the outskirts of galaxies, through interaction and/or merging \citep[][]{Murante2007,Contini2014}. Therefore, studying the main characteristics (i.e. distribution, colors, age, metallicity) of the intra-cluster population constraints the progenitor galaxies and  the processes at work in the formation of the cluster.

Detection and analysis of the intra-cluster component is really challenging, in particular for the ICL, due to  its diffuse nature and very faint surface brightness levels. In the last decade, deep data acquired over wide areas has increasingly contributed to our knowledge of the  ICL properties. This component is mainly concentrated around the brightest galaxies and it is mapped out to several hundred kiloparsec from the center of the clusters \citep{Zibetti2005,Gonzalez2007,Mihos2017}. 
The fraction of stellar envelope plus the ICL around the {  BCG} over the total light in the cluster (including galaxy members) is 10\%-30\%    \citep{Gonzalez2005,Zibetti2005,Seigar2007}. 
The measured colors of the ICL are found {  to be similar to those of the major cluster galaxies}, consistent with the hypothesis that this component could originate from interactions between them  \citep{Zibetti2005}.

In the nearby universe,
deep images of the Virgo cluster ($z\sim0.004$) have revealed several faint ($\mu_V = 26-29$~mag~arcsec$^{-2}$)  streams of ICL between the galaxy members, with the bulk detected close to the BCGs  and colors $B-V=0.7-0.9$~mag \citep{Jan2010,Cap2015,Mihos2017}. The total fraction of  the ICL is $\sim7\%-10\%$. 
This value is smaller than that in massive and evolved clusters and could indicate that the ICL component in Virgo is still growing. 
%
%
%
Also in the Hydra I ($z\sim0.01$)  cluster the ICL is still forming, since remnants of dwarf galaxies are identified
in the outskirts of the central dominating galaxy NGC~3311 \citep{Arnaboldi2012,Hilker2015}.
 

At higher redshift ($z\sim 0.3-0.5$),  the fraction of ICL decreases \citep[5\%-20\%][]{Giallongo2014,Montes2014, Presotto2014}, in agreement with theoretical predictions that suggest a late formation epoch for the intra-cluster population  \citep[$z<1$][]{Murante2007,Contini2014}.

In this paper, we present an unknown ICL region in the core of the Fornax cluster. 
This is the second most massive galaxy concentration within 20 Mpc, after the Virgo cluster, where the mass assembly is still ongoing \citep{Drinkwater2001,Scharf2005}. Most of the bright ($m_B<15$~mag) cluster members in the core are early-type galaxies \citep[ETGs,][]{Ferguson1989}.
The Fornax cluster hosts a vast population of dwarf galaxies and ultra compact galaxies \citep{Munoz2015,Hilker2015a,Schulz2016}, an intra-cluster population of GCs \citep{Bassino2006,Schuberth2010,Dabrusco2016,Cantiello2017} and planetary nebulae \citep{Napolitano2003,McNeil2012}.

In Fornax, a faint ($\mu_g \sim 29-30$~mag~arcsec$^{-2}$) stellar bridge of ICL, $\sim 29$~kpc long, was found between NGC~1399 and NGC~1387  by \citet{Iodice2016}. Based on the color analysis, it could result from the stripping of the outer envelope of NGC1387 on its East side. Such a feature was already detected in the spatial distribution of the blue GCs \citep{Bassino2006,Dabrusco2016}.

In this paper we aim to estimate the total luminosity  and colors of the newly discovered ICL in Fornax (in Sec.~\ref{map}, Sec.~\ref{GC} and Sec.~\ref{phot}) and to discuss its origin (Sec.~\ref{conclu}).


\section{Data: the Fornax Deep Survey with VST} \label{data}

This work is based on the Fornax Deep Survey (FDS) data presented by \citet{Iodice2016}. 
FDS is a multiband ($u$,$g$,$r$ and $i$) imaging survey, obtained with the ESO VLT Survey Telescope (VST), which covers   
$26$~square degrees around the central galaxy NGC~1399 of the Fornax cluster    \citep[see][]{Iodice2016,Iodice2017,Venhola2017}.
{  The limiting $5\sigma$ magnitudes (in AB system) for 1 arcsec$^2$ area are 27.6, 28.5, 28.5 and 27.1 mag in the $u$,$g$,$r$ and $i$ bands, respectively.}

VST is a 2.6-m wide field optical survey telescope, located at Cerro Paranal in Chile \citep{Schipani2012}, 
equipped with the wide field camera ($1 \times1$~degree$^2$) OmegaCam, having a pixel scale of 0.21~arcsec/pixel.

Data reduction was performed by using the VST-Tube pipeline \citep{Grado2012,Cap2015} and the analysis was described in detail by \citet{Iodice2016}. The {\it step-dither} observing strategy, adopted  for the FDS, allows a very accurate estimate of the sky background  \citep{Iodice2016,Iodice2017}. In particular, the average sky frame derived for each observing night takes into account the small contribution to the sky brightness by the smooth components (i.e galactic cirrus, zodiacal light and of the terrestrial airglow)  plus the extragalactic background light.
In the sky-subtracted science frame only a possible differential component could remain,  which contributes to the ``residual fluctuations''  in the background and sets the accuracy of the sky-subtraction step.
 
\begin{figure*}[t]
\includegraphics[scale=0.5]{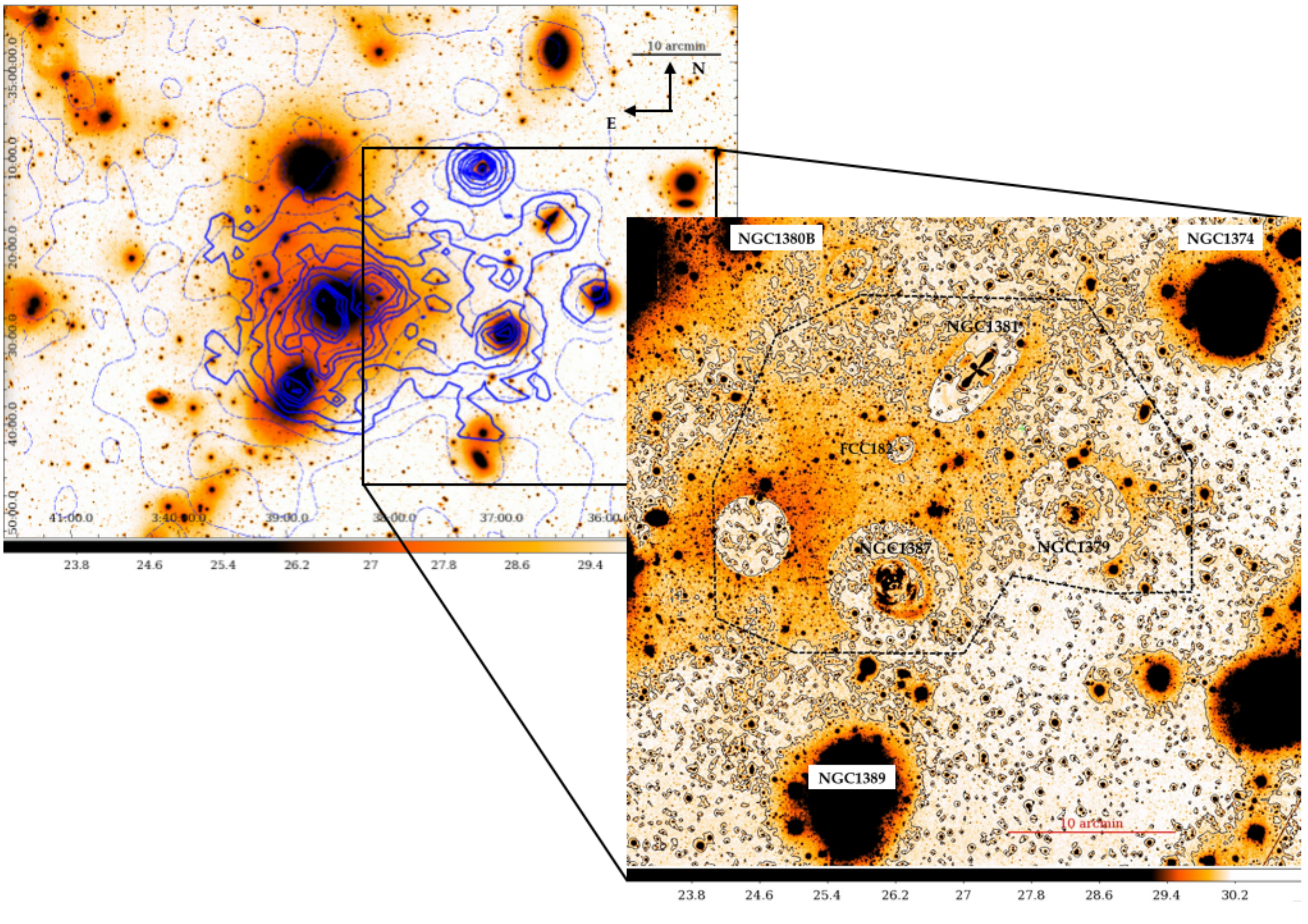}
\caption{\label{mosaic} {\it Left panel -} Central regions ($2 \times 1$~degrees$^2$) of the Fornax cluster in the $r$-band surface brightness levels. The blue contours are the spatial distribution of the blue GCs derived by \citet{Dabrusco2016} (solid lines) and by \citet{Cantiello2017} (dashed lines). {\it Right panel -} Enlarged regions on the West side ($40.9\times42.7$~arcmin $\sim237\times248$~kpc), where the bright galaxies were modelled and subtracted. In this area,  the diffuse ICL is the central over density that fills the intra-cluster space.  The black contours are the surface brightness levels  $\mu_r = 29.8 - 30.02$~mag/arcsec$^2$.  {  The black dashed region  mark the polygon aperture used to  
measure the integrated magnitudes of the ICL (see Sec.~\ref{mag}).}} 
\end{figure*}

In this work we analysed  the core of the Fornax cluster (see left panel of Fig.~\ref{mosaic}) from the $g$ and $r$-band mosaics. We
adopt a distance for NGC~1399 of D=19.95 Mpc \citep{Tonry2001}, therefore the image scale is 96.7 parsecs/arcsec.

\section{Map of the diffuse intra-cluster light} \label{map}

\begin{figure}[t]
\includegraphics[scale=0.45]{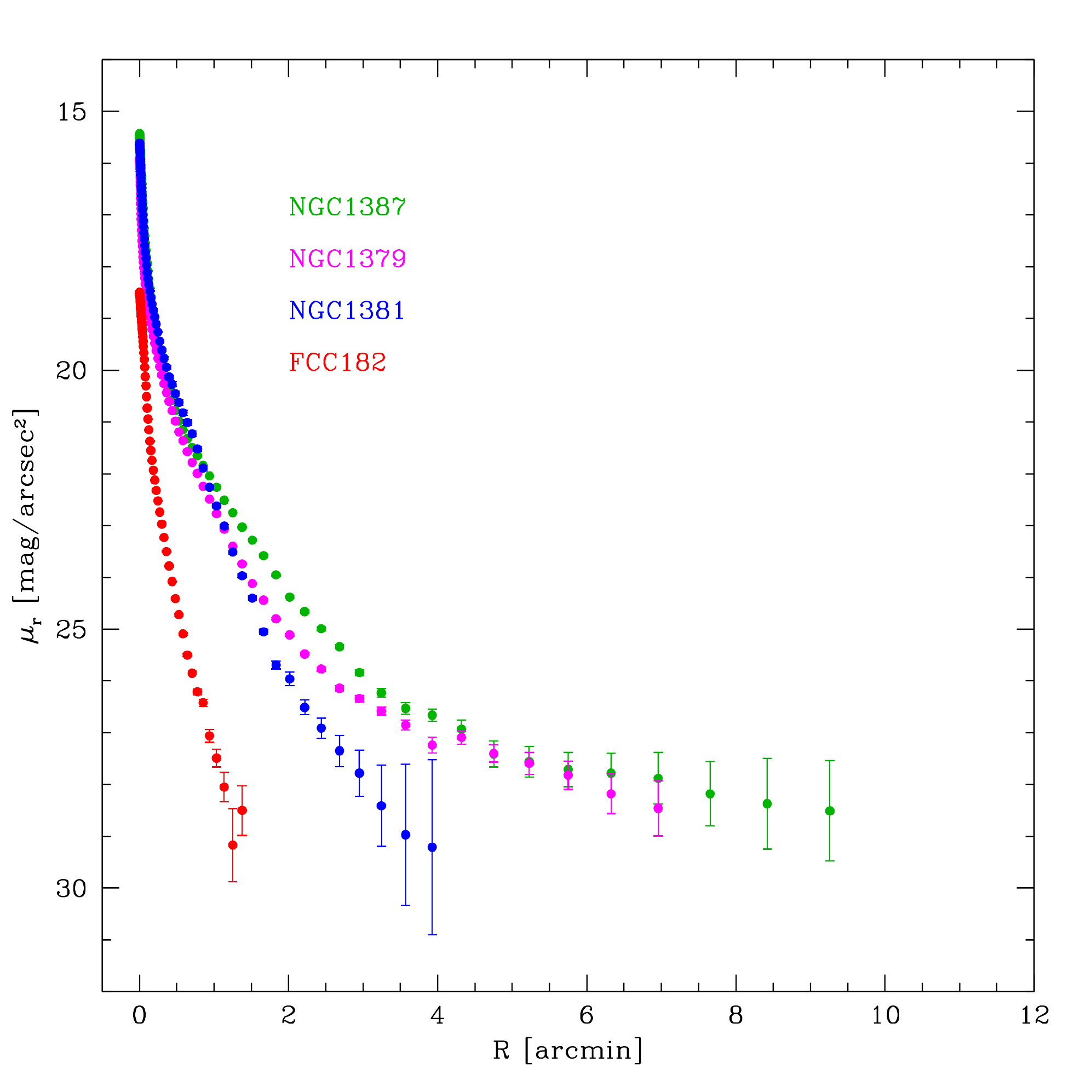}
\caption{\label{prof_ETG}  Azimuthally averaged surface brightness, in the $r$ band, for the three brightest ETGs (NGC~1387, NGC~1379 and NGC~1381) and for dwarf elliptical FCC~182 close to the ICL region.  } 
\end{figure}

On the West side  of the Fornax cluster, close to the NGC~1399 in the core, we detect several {\it patches} of diffuse light in the intra-cluster region (see right panel of Fig.~\ref{mosaic}). 
{  These features are very faint, with a $r$-band surface brightness $\mu_r \simeq 28 - 29$~mag/arcsec$^2$. They  extend out to about 40~arcmin ($\sim230$~kpc) from the center of NGC~1399 and over an area of about 430~arcmin$^2$, in the {  E-W direction} for $-97^\circ \leq P.A. \leq -56^\circ$.
The bulk of the ICL is concentrated between the three bright ETGs in the core, NGC~1387, NGC~1379 and NGC~1381. The dwarf elliptical FCC~182 is completely embedded in this diffuse over-density of light. 

In order to map the ICL and derive the integrated magnitudes and $g-r$ colors, we subtract a two-dimensional (2D) model of the light distribution of all ETGs in the area, including NGC~1399. The right panel of Fig.~\ref{mosaic} shows the residual image.
For each galaxy, the 2D model results from the isophotal analysis made by using the IRAF task ELLIPSE. It takes into account the azimuthally averaged surface brightness distribution as well as the ellipticity and position angle (P.A.) variations  as function of the semi-major axis, down to the average background level in that field. A detailed description of the adopted method is available in  \citet{Iodice2016}, which is a similar approach to that adopted from other studies on the ICL in clusters  \citep[e.g.][]{Krick2007}. 

Fig.~\ref{prof_ETG}  shows the azimuthally averaged surface brightness profiles for the three brightest ETGs (NGC~1387, NGC~1379 and NGC~1381) close to the ICL region, derived by the isophote fit: they extend out to $4-9$~arcmin ($\sim23-50$~kpc) from the galaxy center and map the light down to the faintest galaxy's outskirts ($\mu_r \simeq 29$~mag/arcsec$^2$), i.e. to the region of the stellar halos. 
The 2D model made for each ETG reproduces the whole galaxy light distribution\footnote{  The bright residuals in the inner regions of NGC~1387 and NGC~1381 are due to the presence of substructures, like an inner bar and dust in NGC1387 and a boxy bulge and thick disk in NGC1381. The discussion of such a features, which do not affect the ICL geometry and fraction, is out of the scope of the present work, they will be analysed in detail in a forthcoming paper (Iodice et al. 2017, in preparation).}, 
including the stellar halo (shown in Fig.~\ref{prof_ETG}), and takes into account the variations in ellipticity and P.A. (shown in Fig.~\ref{ETG1} and Fig.~\ref{ETG2}). 
By subtracting the 2D model derived from the above analysis for each galaxy, the residual image clearly shows an excess of light in the intracluster region between the three bright ETGs in the core (see right panel of Fig.~\ref{mosaic}). 
Moreover, the residual image allow us to detect some other patches of ICL  on the NW side, between NGC~1381 and NGC~1380B, between NGC~1381 and NGC~1374, and in the South, between NGC~1399 and NGC~1389 (see right panel of Fig.~\ref{mosaic}). They appear as bridge-like structures connecting the galaxies.

As pointed out in previous studies on ICL \citep{Zibetti2005,Gonzalez2005,Krick2007,Rudick2011,Montes2017},  the separation of the ICL from the extended stellar halo in BCGs is a trick task since the two components tend to merge at the faintest surface brightness levels and, therefore, it is hard to uniquely separate each contribution by deep photometry alone.
The isophotal analysis we performed for the ETGs close to ICL region maps the surface brightness out to the very faintest galaxy outskirts, therefore the 2D model made for each galaxy reproduce it out the region of the stellar halos. Here the fit could include also a small fraction of the ICL, which is mixed with the stellar bounded component. Alternatively, if the outskirts of galaxies deviate in shape from pure ellipses, they would appear as substructures in the residual map and can contribute to the ICL. In Fornax, this is the case for NGC~1379 and NGC~1387. 
The outer isophotes of NGC~1379 are quite extended on the W-NW side, suggesting that this galaxy could have a very asymmetric stellar halo, elongated in the opposite direction of the ICL (see Fig.~\ref{ETG1}, top panel). Something similar was observed for NGC~1387,  whose outer isophotes  extend on the W-NW side (see Fig.~\ref{ETG1}, bottom panel), opposite to the region where the bridge of light toward NGC~1399 is detected  \citep{Iodice2016}. 

In any case, even taking into account the uncertainties in the 2D models of the galaxies at the faintest levels, 
we are confident that the detected ICL in the core of the Fornax cluster does not result from the overlap, along the line of sight, of the galaxies extended stellar halos, but it is a real additional diffuse component in this intracluster region. In fact, we found that  the distance from the center of each object to the bulk of the ICL is between 10-15~arcmin, which is larger than the maximum semi-major axis reached by the surface photometry ($R\sim 4-9$~arcmin, see Fig.~\ref{prof_ETG}).  
}

\subsection{Magnitudes and colors of the ICL} \label{mag}

We derived the  total magnitudes in the $g$ and $r$ bands integrated over an area of $\sim432$~arcmin$^2$, covering the bulk of the ICL.
{  We used the IRAF tasks POLYMARK to trace a polygon including the intra-cluster region between NGC~1387, NGC~1379, NGC~1381 and NGC~1380B, and POLYPHOT to estimate the integrated magnitude inside this area. 
Foreground and background point and extended sources (stars and galaxies) were masked and excluded from the flux estimate.
We account for an average value of the ``local'' residual background level\footnote{ The Òresidual fluctuationsÓ in the sky-subtracted images are the deviations from the  background in the science frame with respect to the average sky frame obtained by the  empty fields close to the target. Therefore, by estimating them, we obtain an estimate on the accuracy of the sky-subtraction step. See \citet{Iodice2016} for a detailed description of this approach.} and the error estimate on the total magnitudes takes into account the scatter in the background level, as also described in \citet{Iodice2016}.}

Magnitudes are corrected for the Galactic extinction  \citep[by using  $A_{g}=0.042$ and $A_{r}=0.029$,][]{Schlegel98} and they are 
$m_g^{ICL}=12.1 \pm 0.3$~mag and $m_r^{ICL}=11.4 \pm0.3$~mag. Therefore, the total luminosity in the $g$ band is $L_g\simeq 8.3 \times 10^{9}$~$L_{\odot}$ \citep[$L_V\simeq 7 \times 10^{9}$~$L_{\odot}$, since $g-V=0.39$ from][]{Fukugita1996}.
%
The fraction of ICL with respect to the total luminosity of NGC~1399 \citep[$L_g\simeq 1.66 \times 10^{11}$~$L_{\odot}$, see][]{Iodice2016} is about  5\%. {  Compared only to the outer extended stellar envelope of NGC~1399 the fraction of the ICL is about  7\%.}

The  $g-r$ integrated color of the ICL is $0.7 \pm 0.4$~mag ($B-V \sim 0.85$~mag). Over the whole extension of the ICL, for $10\leq R\leq 45$~arcmin, we estimate that the color varies in the range $0.6 \leq g-r \leq 1.4$~mag ($ B-V = 0.79 - 1.55$~mag). 
These values are comparable with the $g-r$ colours in the outskirts of the three ETGs close to the ICL region, which are $g-r \sim 0.6$~mag for NGC~1381, $g-r \sim 0.8-1.5$~mag for NGC~1379 and  $g-r \sim 0.8-1$~mag for NGC~1387 (Iodice et al., in preparation). The stellar halo of NGC~1399 also has a similar color, being $g-r \sim 0.8$~mag \citep{Iodice2016}. 

\section{ICL versus intra-cluster GCs}\label{GC}

From the FDS $ugri$ images covering the central $\sim 8.4$ deg$^2$ of the cluster, \citet{Dabrusco2016} traced the spatial distribution of candidate GCs in a region $\sim0.5$~deg$^2$ within the core of the Fornax cluster. The density contours of blue GCs are over-plotted to the light distribution (see Fig.~\ref{mosaic}, left panel). 
\citet{Dabrusco2016} pointed out that the blue GCs density map is elongated on the {  E-W direction}, as also found for the the light distribution \citep{Iodice2016}, and  an over-density of blue GCs is found between  NGC~1381, NGC~1379 and NGC~1380B \citep[see also][]{Bassino2006}. This over-density correlates with the area where the bulk of ICL is detected (see Sec.~\ref{map}). 
Besides, the blue GCs  on NGC~1379 are more concentrated on the East side, falling in the same intra-cluster region. 
There is a strong correlation also between the distribution of blue GCs and the other detected bridge-like structure of ICL: the blue GCs around NGC~1380B appear more elongated toward South as also the diffuse light and an over-density of blue GCs is found into intra-cluster region between NGC~1389 and NGC~1387 (see Fig.~\ref{mosaic}, left panel).

We estimate that the over-density of blue GCs in the NW region, where the total magnitude of the ICL is derived, i.e. $R\leq45$~arcmin (see Sec.~\ref{map}), is $\sim 4$\% of the total population of blue GCs {  of NGC~1399 alone, so all GCs that should be associated to the halos of other galaxies in the core are not counted in}. To do this, we assume that the GCs to not be considered are all candidates inside the outer isophote tracing the stellar halo.

The above results are further confirmed by  the spatial distribution of the blue GCs derived by \citet{Cantiello2017} from the same FDS data (see blue dashed contours in Fig.~\ref{mosaic}). By adopting less stringent selection constraints, the new sample is larger (almost double) and includes all GCs already mapped by \citet{Dabrusco2016}. By using the new GC catalog, the over-density of blue GCs in the NW region of ICL increases to $\sim 7$\%. Both estimates are comparable with the fraction of ICL ($\sim5\%$). 

The correspondence of the spatial and fraction of the over-densities of blue GCs with the newly detected  ICL over different patches is a strong indication for intra-cluster stellar populations in the core of the Fornax cluster.



\section{Radial distribution of the intra-cluster components} \label{phot}

\begin{figure*}[t]
\includegraphics[scale=0.47]{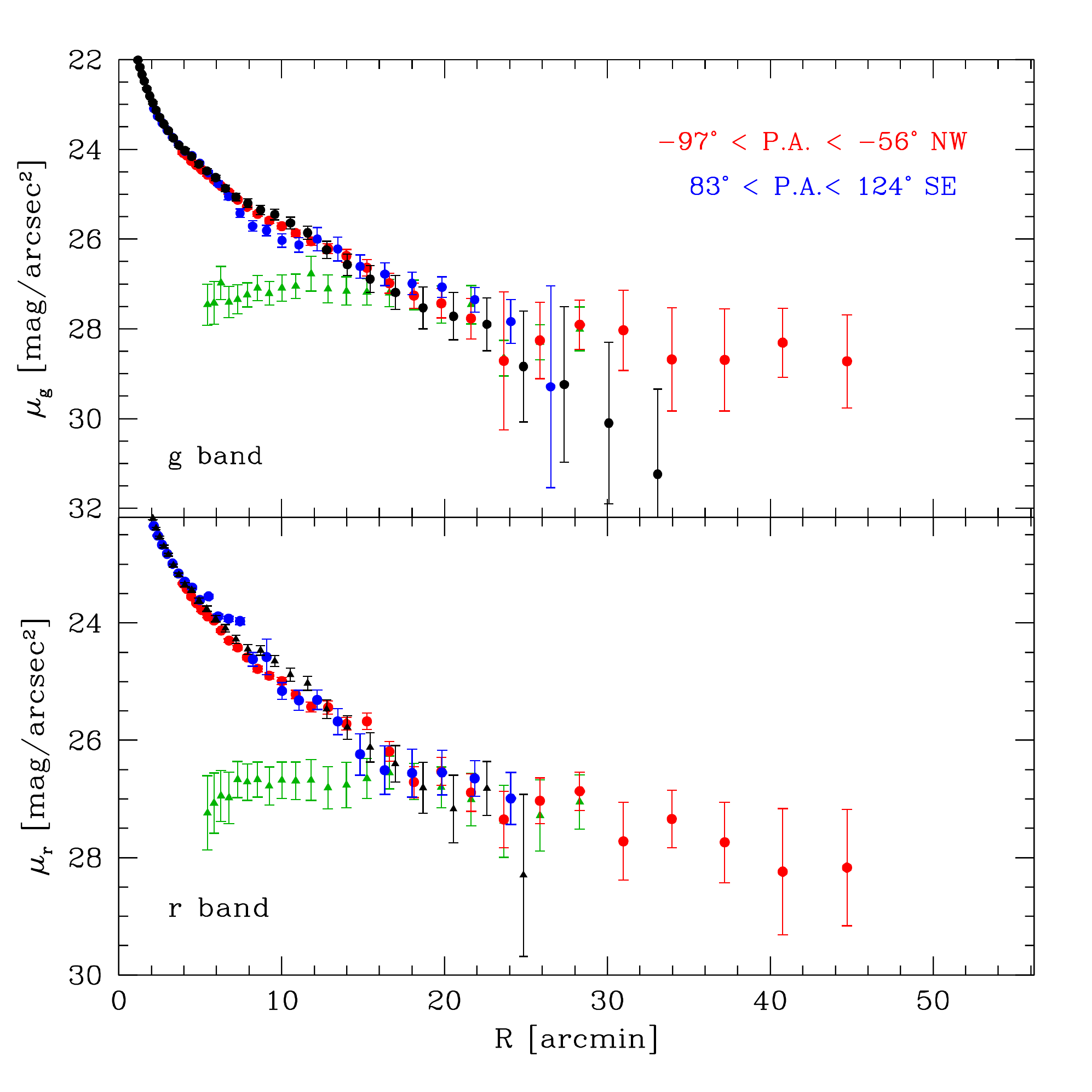}
\includegraphics[scale=0.47]{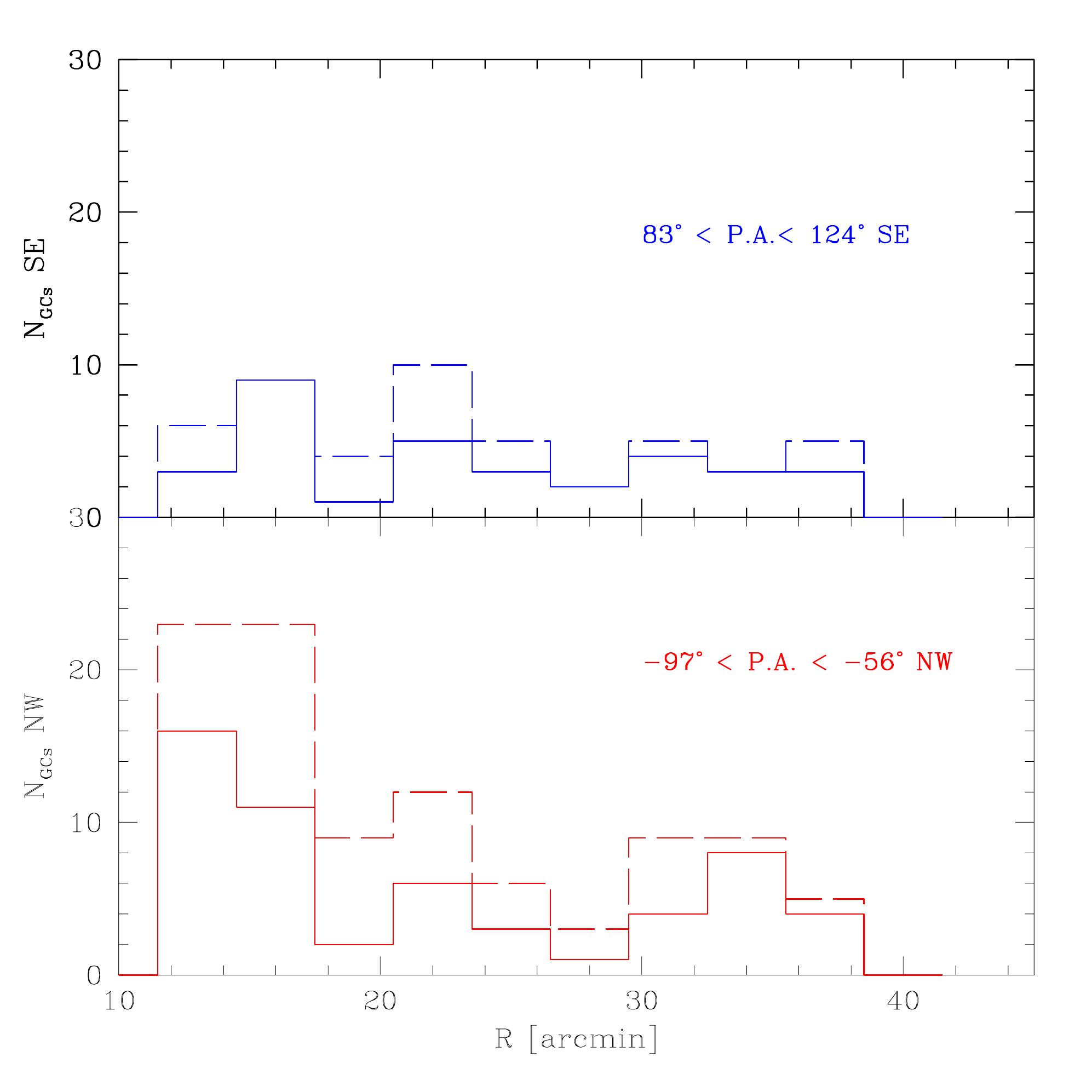}
\caption{\label{prof_ICL}  {\it Left panel} - SB profiles in the $g$  (upper panels) and $r$ bands (bottom panels)  extracted in different areas around NGC~1399. Red circles: SB for $-97^\circ \leq P.A. \leq -56^\circ$, which covers the ICL region. Blue circles: SB for $83^\circ \leq P.A. \leq 124^\circ$, which is the opposite region to the ICL with respect to NGC~1399. Black circles: azymuthally-averaged SB profiles derived by \citet{Iodice2016}. Red triangles: SB distribution of the ICL for $-97^\circ \leq P.A. \leq -56^\circ$ derived on the image where the light from NGC~1399 is subtracted. 
{\it Right panel -}  Number of blue GCs inside the NW (red lines, bottom panel) and SW (blue lines, upper panel) regions where the surface brightness profiles are extracted (left panels), for the sample studied by \citet{Dabrusco2016} (solid line) and  \citet{Cantiello2017} (dashed line).}
\end{figure*}

To analyse the radial distribution of the  ICL, we derived the average surface brightness (SB) profiles in a cone of 40 degrees wide (for $-97^\circ \leq P.A. \leq -56^\circ$), centered on NGC~1399, from {\it i)} the residual $g$ and $r$-band images, where ETGs in this area are subtracted except NGC~1399, and {\it ii)} from the same images {  where} also NGC~1399 is modelled and subtracted.
In both cases, all bright foreground stars and background objects were accurately masked. 
We also derived SB profiles on the SE side, for $83^\circ \leq P.A. \leq 124^\circ$, which is the opposite region to the ICL with respect to NGC~1399.
Results are shown in Fig.~\ref{prof_ICL}. 
On the NW side, the SB extends out to 45 arcmin ($\sim260$~kpc) from the centre of the Fornax cluster. 
Light from the stellar envelope of NGC~1399  \citep[$R\geq10$~arcmin][]{Iodice2016} dominates the SB profiles out to $R\sim20$~arcmin, where  all profiles, in both bands, are consistent with each others.
At larger radii, the $g$ and $r$ light profiles in NW show an excess of light ($\mu\sim28-29$~mag/arcsec$^2$).
In the $g$ band, this is about one magnitude brighter  than both the azimuthally-averaged values and that on the SE side at the same radius. 
In the $r$ band, extralight is detected where the azimuthally-averaged profile in the SE cone have no significant detection. 

Asymmetries in the light distributions with respect to the average were already discussed in  \citet{Iodice2016}. The light profiles extracted along several position angles show an excess of light on the West side of NGC~1399, where the bulk of ICL is found. The most significant intra-cluster region is that in between NGC~1387, NGC~1379 and NGC~1381 ($15 \le R \le 30 $~arcmin),  at P.A.$=98$~degrees West,  where $27 \le \mu_g \le
28$~mag~arcsec$^{-2}$, which is two magnitudes brighter than the light at the corresponding radii on the East side \citep[see Fig.~9 of ][]{Iodice2016}.
%



By subtracting also  the light from NGC~1399, we can trace the ICL at smaller radii than those given above, i.e. $10\leq R \leq 30$~arcmin (see the right panel of Fig.~\ref{mosaic}), {  were the it is completely blended by the light in the stellar halo of NGC1399}. 
For  $-97^\circ \leq P.A. \leq -56^\circ$, it has an almost constant profile, with  $ \mu_g \sim 26.6$~mag~arcsec$^{-2}$ and $ \mu_r \sim 27.2$~mag~arcsec$^{-2}$, showing a maximum for  $10\leq R \leq 20$~arcmin {  (see the left panel Fig.~\ref{prof_ICL})}. This region corresponds to the bulk of the ICL (see Sec.~\ref{map}). On the SE  side ($83^\circ \leq P.A. \leq 124^\circ$) there is no detection above the background level.


Similar to the analysis of the SB profiles, we estimated the number of blue GCs inside the same areas where SB profiles are derived (i.e. NW $-97^\circ \leq P.A. \leq -56^\circ$ and SE $83^\circ \leq P.A. \leq 124^\circ$ cones). The GCs were selected in each SB level as function of the radius (see right panel of Fig.~\ref{prof_ICL}). 
This was done for both samples of blue GCs, by \citet{Dabrusco2016} and \citet{Cantiello2017}.
On average, the blue GC distribution is almost constant with radius on the SE, while there are two peaks on the NW: one at $10\leq R \leq 20$~arcmin, {  where the bulk of ICL is detected} and another for $R\geq 30$~arcmin where we observe an extra-light from the SB profiles.

\section{Discussion: what is the origin of the intra-cluster population?} \label{conclu}

{  
In the core of the Fornax cluster, on the West side of NGC~1399, we detected a previously unknown region of ICL (see right panel of Fig.~\ref{mosaic}). The bulk of ICL is between the three bright ETGs in the core, NGC~1387, NGC~1379 and NGC~1381, at $10\leq R \leq40$~arcmin ($\sim 58 - 230$~kpc) from the center of NGC~1399. 
The ICL  is therefore confined in a small area ($\sim1$\%) compared to the total extension of the Fornax cluster.

One of the major results of this work is the spatial coincidence and fraction of the over-densities of blue GCs with the newly detected ICL over different patches. We found that the ICL is  the counterpart in the diffuse light of the over-density in the blue intra-cluster GCs detected by \citet{Dabrusco2016}. The ICL  in this region of the cluster amounts to $\sim5$\% of the total light of the brightest cluster member NGC~1399. This turns also to be consistent with the fraction of blue GCs ($\sim4-6$\%) with respect to the total population of blue GCs,  in the same area. 
The above findings are a strong indication for intra-cluster stellar populations in the core of the Fornax cluster.

The ICL in the core of the Fornax cluster shows similar properties  to the ICL detected in other clusters of galaxies. Its  total luminosity in the V band is $L_V\simeq 7 \times 10^{9}$~$L_{\odot}$ (see Sec.~\ref{map}). It  is comparable with the total luminosity of  several streams of ICL found in the Virgo cluster, around the BCGs, which are in the range $ 2.3 \leq L_V \leq 5.6 \times 10^{9}$~$L_{\odot}$ \citep{Jan2010,Mihos2017}.
The ICL integrated color is $g-r\sim 0.7$~mag (see Sec.~\ref{map}), corresponding roughly to $B-V \sim 0.85$~mag. This is comparable to the $B-V$ colors derived for the ICL streams in the Virgo cluster \citep{Mihos2017}, which are in the range $0.7-0.9$~mag, and it is also consistent with the $g-r=0.68$~mag found for the ICL in Abell cluster A2744 from \citet{Montes2014}. 
From SB profiles, the ICL colors for $10\leq R\leq 45$~arcmin are in the range $0.6 \leq g-r \leq 1.4$~mag ($ B-V = 0.79 - 1.55$~mag). These  are consistent with $g-r \simeq 1-1.2$~mag found for the ICL in clusters from \citet{Zibetti2005}.

Theoretical studies on mass assembly and ICL formation indicate that the morphology and evolution over  time strongly depend on the processes at work and on the environment. In particular, the main mechanism able to produce the intra-cluster population is the stripping of material from galaxies, which could happen during the initial collapse of the cluster \citep{Merritt1984} or induced by the cluster potential \citep{Byrd1990,Gnedin2003} and/or  from the high-speed encounters between cluster members \citep{Mihos2004,Rudick2006}. During the cluster evolution, all these processes can contribute to the formation of the ICL at different epochs and/or in different regions of the cluster.
Using N-body simulations \citet{Rudick2009} have modelled the ICL formation and  studied the morphology and evolution of this component. Their main results suggest that {\it i)} the morphology depends  on the mechanism that has produced  the ICL, i.e. long streams and tails form during the merging of galaxies, while a close encounter of a galaxy with the cD tends to form shapeless structures like shells or plumes, and  {\it ii)} the evolution is a function of the cluster potential, i.e., once the ICL is formed, its decay time is $\sim1.5$ times the dynamical time in the cluster, before fading into a smooth component. This means that in evolved clusters of galaxies the ICL is in a more diffuse form within the cluster, while, in an early stage formation, the  ICL  appears as long and linear streams.

The qualitative model predictions can be contrasted to our observational findings to constraint the formation history of the intra-cluster component (light and GCs) in the core of the Fornax cluster. 
The ``observables'' are: {\it i)} the morphology of the ICL and {\it ii)}  its location in the cluster. The ICL in Fornax is in the form of faint and diffuse patches, concentrated between the cD (NGC~1399) and  the three bright ETGs, NGC~1387, NGC~1379 and NGC~1381 (see  right panel of Fig.~\ref{mosaic}), for $R \leq 58 - 230$~kpc,
which is within the cluster core and well inside the the virial radius ($R_{vir} \sim 700$~kpc). Compared with the simulations by \citet{Rudick2009}, the above properties of the ICL are consistent with the scenario where this component formed by stripped material from the outskirts of a galaxy in a close passage with the cD. If any stream formed in the core of the cluster it was destroyed within few Gyrs by the strong tidal field of the cluster in this region. Therefore, as Fornax is likely to be  dynamically evolved cluster one would expect to rather see a smooth ICL component than distinct streams in its core. Our findings are consistent with this expectation.

The bright galaxy NGC~1404, close in projection to NGC~1399 on the SE (see  left panel of Fig.~\ref{mosaic}), but with a velocity difference of $\sim500$ km/s to NGC 1399, could have contributed to form such a region of intra-cluster population with stripped material from its outskirt in passages close to NGC 1399 \citep{Bekki2003}.
But also the other ETGs galaxies close to the ICL regions, as NGC~1381, which has a velocity difference of $\sim 300$~km/s far ($V\sim 1724$~km/s) to NGC~1399 ($V\sim 1425$~km/s), and NGC~1379 ($V\sim 1324$~km/s) and NGC~1387 ($V\sim 1302$~km/s) could have contributed to the growth of the ICL via tidal interactions with the cluster potential.
Such a process was invoked to explain the origin of the stellar bridge connecting NGC~1399 and NGC~1387 \citep{Iodice2016}, which is part of this ICL region. This scenario is further supported by finding that two of the major galaxies in the ICL region (NGC~1387 and NGC1379) have a lopsided stellar halo (see Fig.~\ref{ETG1}), indicating 
gravitational displacements via tidal interaction,  with average colors $0.6 \leq g-r \leq 1.5$~mag comparable with the average value estimated for the ICL in Fornax (see Sec.~\ref{phot}). 

{  Even if the analysis performed and discussed in this paper suggests that the intra-cluster population (GCs and diffuse light) in the core of the Fornax cluster could origin from the outskirts of the galaxies in this region, for the sake of completeness we mention below any other possible 
formation process for this component.

A fraction of the ICL population in this region of the cluster could also come from lower-mass dwarf galaxies, tidally disrupted in the potential well of the massive galaxies.  This is supported by a recent study from \citet{Venhola2017} that found a drop in the number density of LSB galaxies  at cluster-centric distances smaller than $\sim180$~kpc inward.

The blue GCs found in the region of the ICL have  {  typical} $g-r\sim0.5$~mag \citep{Dabrusco2016} bluer  than the diffuse light. This is expected, since specific {  frequencies of the blue (old and metal-poor) GCs  are higher} in the outer halos of galaxies, whereas the stellar population might be, on average, redder and more metal-rich \citep[][and reference therein]{Lamers2017}. This would also indicate that GCs might come from less massive and hence less metal-rich disrupted LSB galaxies, since the specific frequency of GCs strongly increases toward lower-luminosity dwarf galaxies \citep{Peng2008}.
}

As conclusive remark, we stress that the spatial coincidence of the ICL with the GCs found in the core of the Fornax cluster  turned out a really powerful tool to trace the intra-cluster baryons and, therefore, to study the formation and evolution of this component. If the same analysis is performed on other clusters of galaxies and the same coincidence is still found, this would suggest that 2D  maps of GCs could be considered as a valid alternative to the light distribution maps to study the intra-cluster population. This perspective should be more properly considered in the era of the new generation of large telescopes and adaptive optics, since they could provide an easier detection of the intra-cluster compact sources rather than of the diffuse light, which requires tricky tasks for data reduction and analysis. 
}



\acknowledgments
This work is based on visitor mode observations taken at the ESO La Silla Paranal Observatory within the VST GTO Program ID 094.B-0496(A). 
{   The authors wish to thank the anonymous referee for his/her comments and suggestions
    that allowed us to greatly improve  the paper. }
Authors acknowledges financial support from the INAF VST funds and  wish to thank ESO for the financial contribution given for the visitor mode runs at the ESO La Silla Paranal Observatory. 
The authors acknowledges financial support from the European UnionÕs Horizon 2020 research and innovation programme under the Marie Skodowska-Curie
grant agreement No 721463 to the SUNDIAL ITN network.
NRN and EI received support within PRIN INAF 2014 "Fornax Cluster Imaging and Spectroscopic Deep Survey".
GvdV acknowledges partial support from Sonderforschungsbereich SFB 881 "The Milky Way System" (subprojects A7 and A8) funded by the German Research Foundation.

\appendix

\section{Images of the galaxies close to the ICL region}

\begin{figure*}[b]
\includegraphics[scale=0.47]{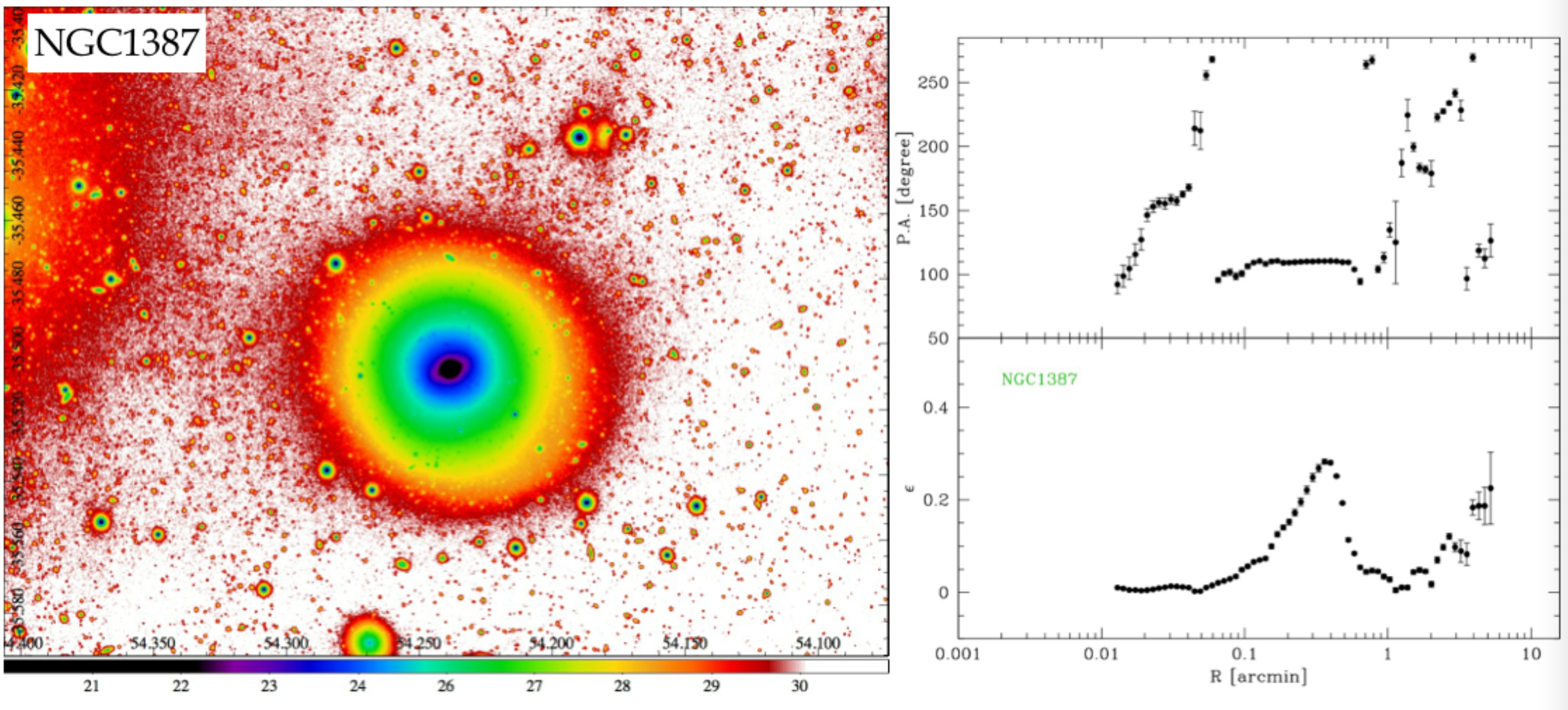}
\includegraphics[scale=0.47]{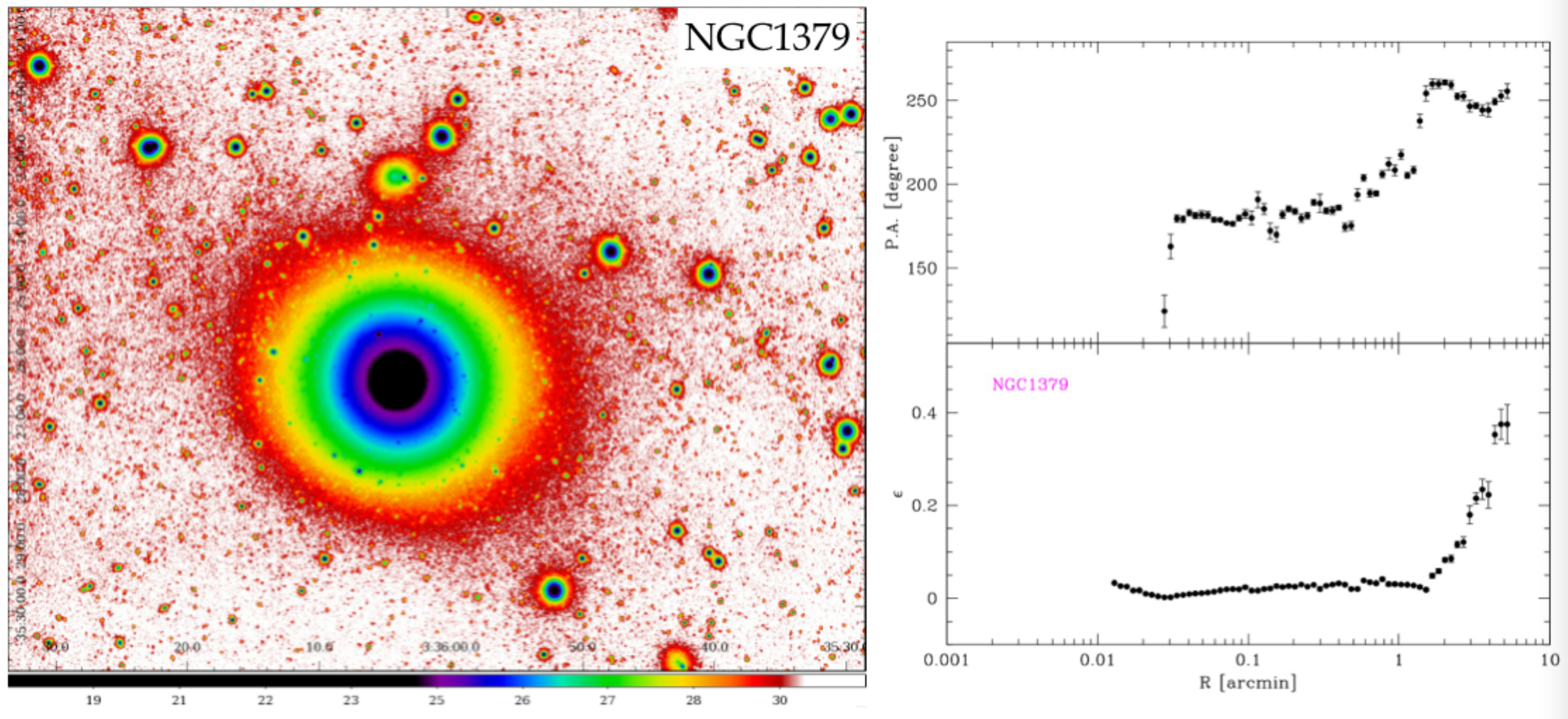}
\caption{\label{ETG1}   Extracted images from the $r$-band VST mosaic of the core of the Fornax cluster, shown in Fig.~\ref{mosaic}, around the galaxy NGC~1387 (top-left panel) and NGC~1379 (bottom-left panel). Images are shown in surface brightness levels reported in the colorbar on the bottom. On the right panels  the ellipticity (top) and P.A. (bottom) profiles  are shown, as function of the semi-major axis, derived by fitting the isophotes in the $r$ bands images.  }
\end{figure*}

\begin{figure*}[t]
\includegraphics[scale=0.47]{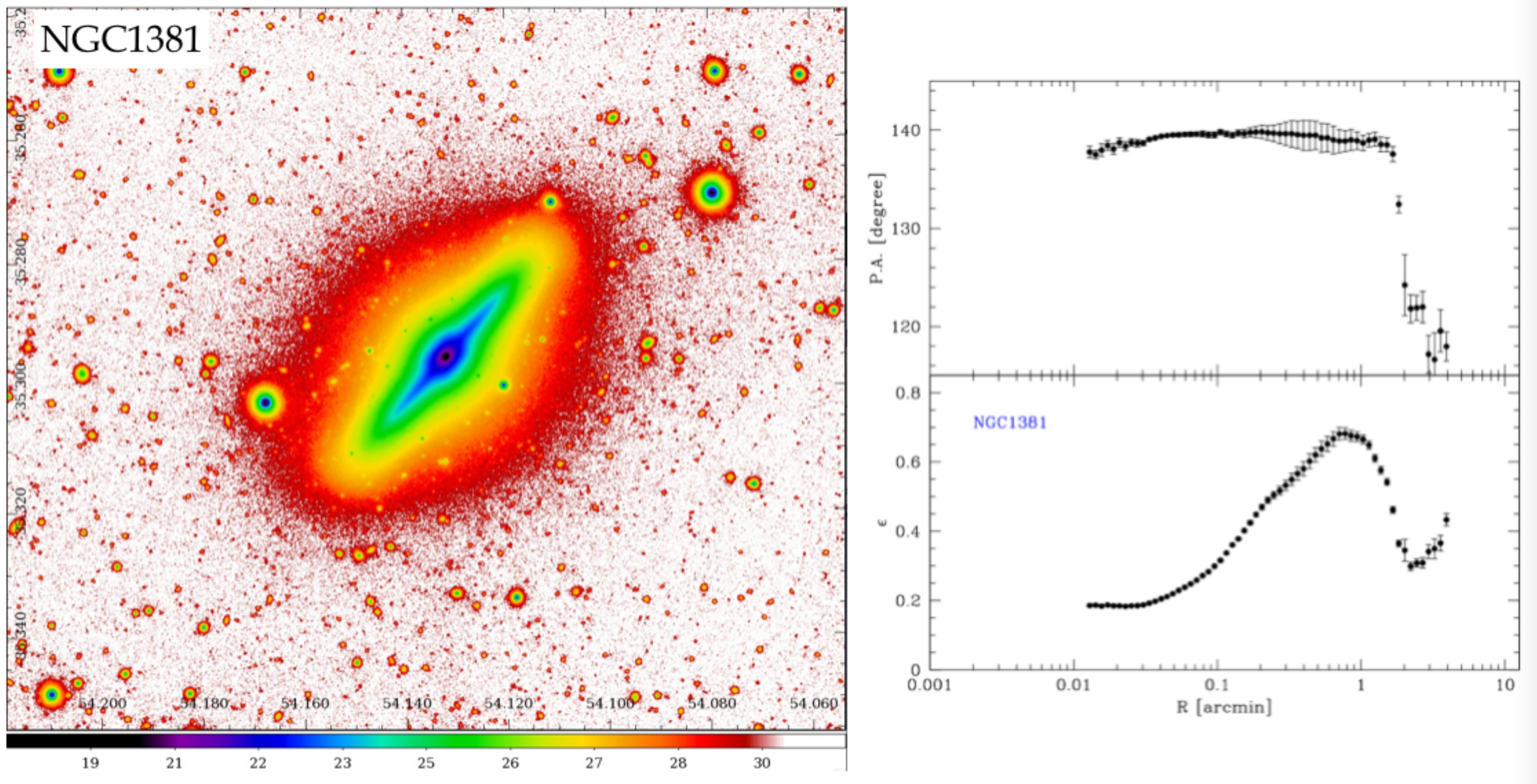}
\includegraphics[scale=0.47]{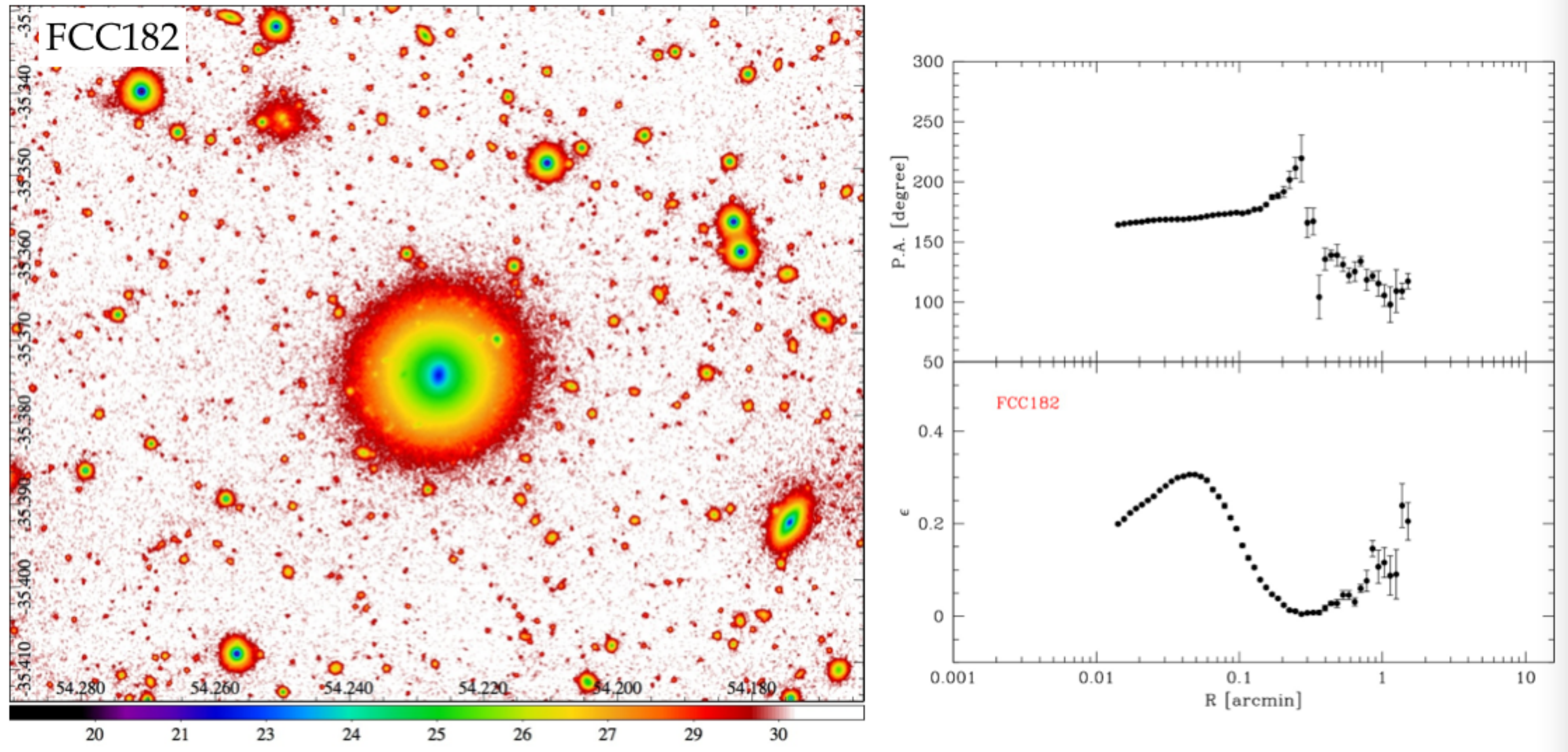}
\caption{\label{ETG2} Same as Fig.~\ref{ETG1} for the S0 galaxy NGC~1381 (top panels) and for the dwarf elliptical FCC182 (bottom panels).   }
\end{figure*}

In this section we show the images of the three bright ETGs in the core of the Fornax cluster, NGC~1387, NGC~1379 (see Fig.~\ref{ETG1}) and NGC~1381 (Fig.~\ref{ETG2}, top panel), and of the dwarf elliptical FCC~182 (Fig.~\ref{ETG2}, bottom panel), which is completely embedded in this diffuse over-density of light (see Fig.~\ref{mosaic}). These are in surface brightness levels in the $r$ band and were extracted from the whole VST mosaic shown in Fig.~\ref{mosaic}. In the right panels of  Fig.~\ref{ETG1} and Fig.~\ref{ETG2} we show the ellipticity and P.A. profiles, as function of the semi-major axis, derived by fitting the isophotes in the $r$ bands images.

\clearpage

\end{document}